\documentclass[apl,twocolumn,showpacs,amsmath,letter]{revtex4}
\usepackage{graphicx}

\newcommand{\Journal}[4]{#1 \textbf{#2}, #3 (#4)}

\begin{document}

\title{Effect of Interlayer Coupling on Current-Assisted Magnetization Switching in Nanopillars}

\author{S. Urazhdin}
\author{W. P. Pratt Jr.}
\author{J. Bass}
\affiliation{Department of Physics and Astronomy, Center for Fundamental Materials Research,
and Center for Sensor Materials, Michigan State University, East Lansing, MI 48824-2320}


\begin{abstract}
We show that dipole-field induced antiferromagnetic coupling, or
RKKY ferromagnetic coupling, between Co layers can strongly affect
the low magnetic field switching behavior of Co/Cu/Co nanopillars.
Whereas current-assisted switching at low fields in uncoupled
nanopillars is always hysteretic, strong coupling of either kind
can change the switching to non-hysteretic (reversible).  These
differences can be understood with a simple picture of
current-assisted thermal activation over a barrier.
\end{abstract}

\maketitle

Current-assisted switching in magnetic nanopillars has lately been
receiving much theoretical~\cite{slonczewski,berger,waintal,
sun,heide,zhang,stiles}  and
experimental~\cite{cornellorig,cornellapl,grollier,cornelltemp,cornellquant,wegrowe,sun2}
attention, both to understand the basic physics and because of its
potential for technology.  Most of the experimental data have been
taken on Co/Cu/Co nanopillars that are either magnetically
uncoupled or antiferromagnetically (AF) coupled.   Initial studies
focused upon the similarities in behavior of the two cases.
However, we recently showed~\cite{urazhdinapl} that at low
magnetic field H the dependencies on driving current I of
uncoupled and AF coupled nanopillars could be very different:
hysteretic for uncoupled samples, but non-hysteretic (reversible)
and characterized by telegraph noise for AF coupled ones. In this
short paper we provide additional data showing these differences
for uncoupled and AF coupled nanopillars, and extend the study to
show that low field reversible behavior and telegraph noise occur
also in ferromagnetically (F) coupled nanopillars.  We show that
these differences in behavior can be understood using a simple
model of current-assisted thermal activation.

Our samples have the form [Cu(80)/Co(20)/Cu(10 or 2.6)/
Co(2.5)/Cu(5)/Au(200)], where thicknesses are in nm. Sample
preparation procedures are described elsewhere~\cite{urazhdinapl}.
In uncoupled samples, only the top Co(2.5) and most of the middle
Cu layer were ion milled to nanopillar size ($\approx$70 nm x 130
nm), leaving the rest of that Cu layer and the bottom Co(20)-layer
extended. This geometry minimizes dipolar coupling between the two
Co layers.  AF coupling was achieved by milling about halfway
through the Co(20) layer, thereby generating dipolar coupling
between the two patterned Co layers. F coupling was achieved by
reducing the Cu thickness to Cu(2.6), near the third RKKY
magnetoresistance (MR) minimum~\cite{parkin}. There were
significant variations in coupling strength among both AF and F
coupled samples, presumably due to sample shape variation and
roughness of magnetic interfaces.

\begin{figure}
\includegraphics[scale=0.41]{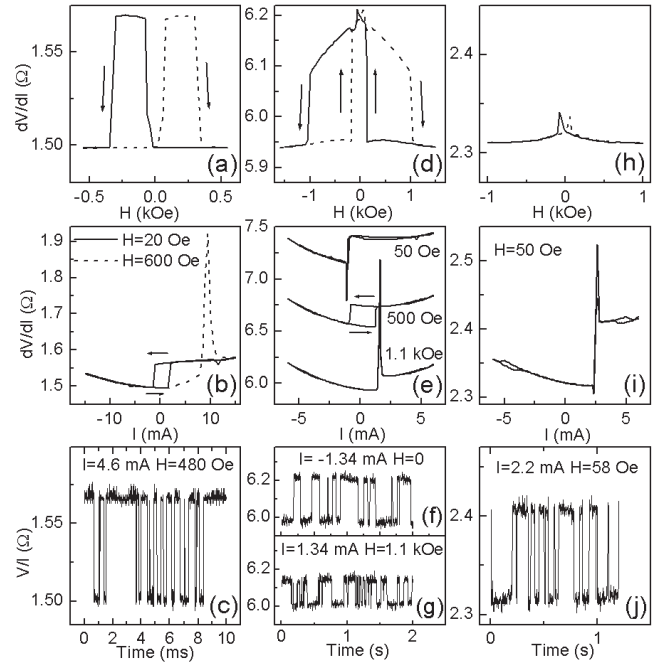}
\caption{\label{fig1} Results for uncoupled (a-c), AF-coupled
(d-g), and F-coupled samples (h-j). In the hysteretic plots,
arrows show the scan direction. (a,d,h) dV/dI {\it vs.} H at
$I=0$. (b,e,i) dV/dI {\it vs.} I at the specified values of H.
(c,f,g,j) Time-resolved measurements of R=V/I at the specified
values of H, I.}
\end{figure}

Differential resistances, dV/dI, were measured with four probes
and lock-in detection, adding an ac current of amplitude 20~$\mu$A
at 8 kHz to the dc current I. Positive current flowed from the
extended to the fully patterned Co layer. Fig.~\ref{fig1}
summarizes the differences between uncoupled (left), AF-coupled
(middle), and F-coupled (right) samples. At I=0, the uncoupled
(Fig.~\ref{fig1}(a)) and AF-coupled (Fig.~\ref{fig1}(d)) samples
display the usual changes from a low resistance, high-H  state in
which the magnetizations of  Co layers are aligned parallel (P) to
each other, to a high resistance, low field state where the
magnetizations are aligned antiparallel (AP). In contrast, in
strongly F-coupled samples the magnetizations are simultaneously
reversed at small H to stay in the exchange-favored P state,
yielding only a small feature in MR at I=0 (Fig.~\ref{fig1}(h)).
At large enough $I>0$, F-coupled samples yielded ~5\% MR (with no
hysteresis), similar to the values obtained for uncoupled and
AF-coupled samples at $I=0$. For uncoupled samples, the change in
dV/dI usually occurs in a single step, indicating single domain
switching.   For AF coupled samples, the non-uniform dipolar field
usually leads to a more complex structure. Fig.~\ref{fig1}(b,e,i)
compares the variations of dV/dI with I for the same three
samples.  First, we compare the behaviors at small H=20-50~Oe,
applied to fix the magetization state of the bottom Co layer. The
uncoupled sample (Fig.~\ref{fig1}(b), solid line) shows the
expected asymmetric hysteretic switching~\cite{cornellorig}
between the same values of dV/dI as in Fig.~\ref{fig1}(a). In
contrast, the AF-coupled sample (Fig.~\ref{fig1}(e), top curve)
shows reversible (non-hysteretic) switching at a negative value of
I, and the F-coupled sample (Fig.~\ref{fig1}(i)) shows reversible
switching at a positive value of I. The dashed line in
Fig.~\ref{fig1}(b) shows that at larger H the switching in
uncoupled samples becomes nonhysteretic, characterized by a peak
similar to that in Fig.~\ref{fig1}(i) for an F-coupled
sample. In the AF-coupled sample, the switching becomes hysteretic
at intermediate H (Fig.~\ref{fig1}(e), middle curve), and
nonhysteretic again at large enough H  (Fig.~\ref{fig1}(e), bottom
curve). These curves are similar to those for the uncoupled sample
(Fig.~\ref{fig1}(b)), but offset by the dipolar coupling field.

Time-resolved measurements, performed at I, H, close to the
nonhysteretic switching peaks, are characterized by telegraph
noise switching between the AP and P states.
Figs.~\ref{fig1}(c,f,g,j) show examples for the uncoupled sample
and the AF-coupled sample at high H, and both coupled samples at
small H. Note that, at identical I of opposite signs, the average
telegraph noise periods in the AF-coupled sample are similar.

\begin{figure}
\includegraphics[scale=0.4]{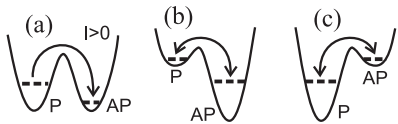}
\caption{\label{fig2} Schematics of current driven switching.
Dashed lines indicate the effective magnetic temperature. (a)
P$\to$AP switching at $I>0$ in the hysteretic regime, at small H
in uncoupled, and intermediate H in AF-coupled samples. (b)
Telegraph noise at small H and $I<0$ in AF-coupled samples. (c)
Telegraph noise at large H in uncoupled and AF-coupled samples,
and at small H in F-coupled samples.}
\end{figure}

The similar high-H behaviors, and different low-H behaviors among
the samples shown in Fig.~\ref{fig1} can be understood using a
simple model of current-assisted thermal excitation over a
magnetic barrier separating the AP and P states~\cite{urazhdin}.
We do not rule out alternative interpretations that may lead to
similar results~\cite{zhang2}. Our model assumes that large enough
current $I>0$($I<0$) generates magnetic excitations in the P(AP)
state, that are described by current-dependent effective magnetic
temperatures $T^{P(AP)}_m$. The magnetic barrier between the P and
AP states is assumed to vary only through the temperature
dependence of magnetization. Fig.~\ref{fig2} shows schematics for
the different cases of interest. Fig.~\ref{fig2}(a) is for
hysteretic transitions, which occur both in uncoupled samples at
low H or in AF-coupled samples when H balances the dipolar
coupling field. The P$\to$AP and AP$\to$P barriers are the same,
but $I>0$ results in $T^P_m>T^{AP}_m$, leading to P$\to$AP
switching. Similarly, at $I<0$, $T^{AP}_m>T^P_m$, leading to
AP$\to$P switching. Fig.~\ref{fig2}(b) is for AF-coupling at low
H. Because the P$\to$AP barrier is small, the P$\to$AP transition
is thermally activated at $I=0$. Large enough $I<0$ (causing
$T^{AP}_m>T^P_m$) also activates the reverse AP$\to$P transition,
leading to telegraph noise. Fig.~\ref{fig2}(c) shows that, at
large enough H, the P$\to$AP barrier is larger than AP$\to$P
barrier in all the samples. In this case, both transitions are
thermally activated at large enough $I>0$.

We acknowledge helpful discussions with Norman O. Birge and support from the
MSU CFMR, CSM, the MSU Keck
Microfabrication facility, the NSF through Grants DMR 02-02476 and
98-09688, and Seagate Technology.

\end{document}